# Trade Openness, Tariffs and Economic Growth: An Empirical Study from Countries of G-20

S M Toufiqul Huq Sowrov

## Abstract

International trade has been in the forefront of economic development and growth debates. Trade openness, its definition, scope, and impacts have also been studied numerously. Tariff has been dubbed as negative influencer of economic growth as per conventional wisdom and most empirical studies. This paper empirically examines relationships among trade openness as trade share to GDP, import tariff rate and economic growth. Panel dataset of 11 G-20 member countries were selected for the study. Results found a positively significant correlation between trade openness and economic growth. Tariff has negatively significant correlation with economic growth in lagged model. OLS and panel data fixed-effects regression were employed to carry out the regression analysis. To deal with endogeneity in trade openness variable, a 1-year lag regression technique was conducted. Results are robust and significant. Policy recommendation suggests country specific trade opening and tariff relaxation.

# List of Contents

Contents

# CHAPTER I: INTRODUCTION

## 1.1 Introduction

Economy and economic activities have been in the heart of human civilization since the beginning. To pursue the ultimate economic growth and betterment, mankind has been developing various economic means. Trade, later international trade, is one of the most important aspects of those economic means which brought immense wealth to nations. Also, there are various examples of losing stories from international trade. Apart from those strains, trade has been regarded as the best tool to alleviate and eradicate poverty, create new wealth, and foster economic growth. Therefore, it has been under study and constant examinations by economists to understand the said correlation between trade and economic growth.

International trade has been an integral part of economic thought since antiquity. International trade, henceforth IT, in the form of export and import has remained a great concern for the countrys' policymakers. The connection between international trade and economic growth has been loitered as a contested topic from the perspectives of a few main issues like export led growth and import substitution. Mercantilist theories see benefits in export surpluses, thus advocates for export over import, trade barriers. Therefore, tariffs and trade barriers are frequently imposed on imports and, seldom, exports. Classical economists like Adam Smith who saw benefits in unhindered and barrier free trade as he argued that trade growth increases specialization in his famous literature work. This specialization leads to economies of scale and greater markets for small domestic economies. David Ricardo's Theory of Comparative Advantange refers to economy's ability to produce goods and services at a lower opportunity cost than that of trade partners. The Ricardian model suggests economies can benefit from specialization in sectors where they have comparative advantages. The Heckler-Ohlin model of IT shows that specialization based on factor endowment can lead to static productivity and economic benefits in increased international trade. Further, New growth theory has provided important insights into an understanding of the relationship between trade and growth. These models exhibit international trade can promote economic growth through employment generation, income distribution, technology spill over and external stimulation. Countries can reap welfare through full employment, better allocation of resources as trade creation gains that arise from increased volume of trade, though gains from trade are not equally distributed. Trade has played its part in efficient allocation of resources in countries and among countries, and being the engine of global growth, however, not all countries share the growth or enjoy the benefits it brings. Developing countries have a much smaller share in global trade values and volumes whereas the developed world enjoys the lion share. Developing countries chiefly produce and export raw and primary materials. Even though trade barriers are falling, the developed world still retains some protection to protect their own markets from imports. In this paper, I will empirically test relationships among economic growth, trade growth or openness and trade tariff (percentage of total revenue). I will address this issue later.

On the other side of the coin, classical economists acknowledged the role of open trade in economic growth, however, the neoclassical economists have given small emphasis on role of trade. Neoclassical economists put importance on endogenous factors of economic growth such as domestic capital formulation, labor, science and technology and human capital. Weaknesses in this model, however, led to works which modified these shortcomings. Endogenous growth models, and subsequent new growth theory, have focused on different variables and look insight of nexus between growth and trade. All these lead to assumptions that trade enables access for a country to technological knowledge and allows manufacturers and growers to access bigger markets through integration with the world economy. Thus, trade boosts countrys' productivity, economic growth. But inferring to conclusion from these studies and theories can be inconclusive, for example, effects of tariffs on trade openness and economic growth is ambiguous and not settled.

Most of the works on trade, tariff and growth nexus find out 4 (four) kinds of relations between trade and growth. Those are: 1) Both export-import-led growth, 2) growth-led trade, 3) both way relations (bidirectional), 4) no causal relations. It is suggested that gains from trade largely depend on proper institutions in place. Otherwise, trade gains, reform and other structural adjustments will bear no fruit. If balance of trade causes balance of payments difficulties, and it is not self-correcting, then trade gains can effectively be canceled out. Thus, special, and careful consideration should be placed on the pace of trade liberalization and protection from external trade.

In terms of protection from external trade, tariffs are the most favored measures among the states. It is easy to implement and one of the fastest ways to collect revenue. It is also a bone of contention in trade negotiations. It has been shown in literal works of the World Bank Group (WB) and International Monetary Fund (IMF) that tariffs have positive correlation with corruption in public sector. Given the importance of its effects on free trade, it is still under studied.

The study of both trade and tariffs with economic growth is particularly important as our world has been seeing frequent imposition of tariffs on trade recently. Popular cases are from the United States particularly. Also, as the United Kingdom has left the European Union, reemergence of tariffs on bilateral trade is real possibilities. The IMF and World Bank are forecasting a global recession in 2020 which is very much probable. During the recession time, it is fundamentally important for states to cushion the shock, take measures and get back to pre-recession economic growth. Therefore, empirical study of factors which construct gross domestic products (GDP) is utmost important. In the days of highly likely global recession, countries will see ever less trading among states. Thus, the countries dependent on trade-led growth will be hottest hard.

**1.2 Motivation**

The motivation behind this study is to revisit the relationships among trade openness, in percentage to total GDP, direct tariff rates on all imported products and economic growth for the 11 countries from G-20 forum. The said relationships are much studied yet concrete conclusion or results to be found. Rationale behind choosing G-20 as sample is countries of G-20 comprise 85% of global GDP and lion share of world's growth rate. Those 11 countries are: Argentina, Australia, Brazil, Canada, France, India, Indonesia, Republic of Korea, Japan, South Africa, and the United States.

The set of these countries has diversity, both developing and developed economies. To answer the question why 11 countries, out of 20 members, because of their necessary data availability. As I have scoured the internet for relevant data, these countries promised something useful I can work on. G-20 consists of both developed and developing economies. 11 of those countries are from 1st and 3rd world countries, thus offer varieties in data. I have discussed further in the Data section. The purpose of the study is to provide statistical and analytical evidence of the impacts trade openness and tariffs have on economic growth of those countries. It is also to find out the causation and correlation between important variables which comprise countrys' GDP.

### 1.2.1 Country overview

Before moving onto the next sections, it would be useful to have a brief sight on the examined countries. As has been mentioned already, the number of countries examined here is 11 (eleven). These countries are members of various inter-governmental organizations (IGO). Even though these countries are members of G-20 and big economies, this set of countries have lower-middle income to high income economies. We will have a brief introduction to these countries, chiefly about important aspects of economies, politics, and development. All the data have been retrieved from the Central Intelligence Agency or CIA's world fact book.

a) **Argentina:**
- Country of South America
- Capital: Buenos Aires
- Year of Independence: 1816 from Spain
- Total area: 2,780,400 sq km
- Population: 45,479,118
- Language: Spanish (official), Italian, English, German, French, indigenous (Mapudungun, Quechua)
- Literacy: 99%
- Unemployment rate: 8.4% (2017 est.)
- Government: Presidential Republic
- GDP (PPP): $922.1 billion (2017 est.)
- GDP (real growth rate): 2.95%
- GDP per capita (PPP): $20,900 (2017 est.)
- Inflation rate (cpi): 25.7% (2017 est.)

b) **Australia:**
- Country of Oceania
- Capital: Canberra
- Year of Independence: 1901 from the United Kingdom
- Total area: 7,741,220 sq km
- Population: 25,466,459 (July 2020 est.)
- Language: English 72.7%, Mandarin 2.5%, Arabic 1.4%, Cantonese 1.2%, Vietnamese 1.2%, Italian 1.2%, Greek 1%, other 14.8%, unspecified 6.5% (2016 est.)

- Literacy: 99%
- Unemployment rate: 5.6% (2017 est.)
- Government: federal parliamentary democracy under a constitutional monarchy
- GDP (PPP): $1.248 trillion (2017 est.)
- GDP (real growth rate): 2.2% (2017 est.)
- GDP per capita (PPP): $50,400 (2017 est.)
- Inflation rate (cpi): 2% (2017 est.)

**c) Brazil:**
- Country of South America
- Capital: Brasilia
- Year of Independence: 1822 from Portugal
- Total area: 2,780,400 sq km
- Population: 211,715,973 (July 2020 est.)
- Language: Portuguese (mostly)
- Literacy: 93.2%
- Unemployment rate: 12.8% (2017 est.)
- Government: Federal Presidential Republic
- GDP (PPP): $3.248 trillion (2017 est.)
- GDP (real growth rate): 1% (2017 est.)
- GDP per capita (PPP): $15,600 (2017 est.)
- Inflation rate (cpi): 3.4% (2017 est.)

**d) Canada:**
- Country of North America
- Capital: Ottawa
- Year of Independence: 1867 (Self-governing dominion from the United Kingdom), 1982- constitution repatriated.
- Total area: 9,984,670 sq km
- Population: 37,694,085 (July 2020 est.)
- Language: English (official) 58.7%, French (official) 22%, Punjabi 1.4%, Italian 1.3%, Spanish 1.3%, German 1.3%, Cantonese 1.2%, Tagalog 1.2%, Arabic 1.1%, other 10.5% (2011 est.)
- Literacy: 99%
- Unemployment rate: 6.3% (2017 est.)
- Government: federal parliamentary democracy under a constitutional monarchy
- GDP (PPP): $1.774 trillion (2017 est.)

- GDP (real growth rate): 3% (2017 est.)
- GDP per capita (PPP): $48,400 (2017 est.)
- Inflation rate (cpi): 1.6% (2017 est.)

**e) France:**
- Country of Europe
- Capital: Paris
- Year of Independence: N/A
- Total area: 643,801 sq km
- Population: 67,848,156 (July 2020 est.)
- Language: French (official)
- Literacy: N/A
- Unemployment rate: 9.4% (2017 est.)
- Government: semi-presidential republic
- GDP (PPP): $2.856 trillion (2017 est.)
- GDP (real growth rate): 2.3% (2017 est.)
- GDP per capita (PPP): $44,100 (2017 est.)
- Inflation rate (cpi): 1.2% (2017 est.)

**f) India:**
- Country of Asia
- Capital: New Delhi
- Year of Independence: 1947 from the United Kingdom
- Total area: 3,287,263 sq km
- Population: 1,326,093,247 (July 2020 est.)
- Language: Hindi 43.6%, Bengali 8%, Marathi 6.9%, Telugu 6.7%, Tamil 5.7%, Gujarati 4.6%, Urdu 4.2%, Kannada 3.6%, Odia 3.1%, Malayalam 2.9%, Punjabi 2.7%, Assamese 1.3%, Maithili 1.1%, other 5.6% (2011 est.)
- Literacy: 74.4%
- Unemployment rate: 8.5% (2017 est.)
- Government: federal parliamentary republic
- GDP (PPP): $9.474 trillion (2017 est.)
- GDP (real growth rate): 6.7% (2017 est.)
- GDP per capita (PPP): $7,200 (2017 est.)
- Inflation rate (cpi): 3.6% (2017 est.)

g) **Indonesia:**
- Country of Asia
- Capital: Jakarta
- Year of Independence: 1945 from the Netherlands
- Total area: 1,904,569 sq km
- Population: 267,026,366 (July 2020 est.)
- Language: Bahasa Indonesia (official, modified form of Malay), English, Dutch
- Literacy: 95.7%
- Unemployment rate: 5.4% (2017 est.)
- Government: presidential republic
- GDP (PPP): $1.015 trillion (2017 est.)
- GDP (real growth rate): 5.1% (2017 est.)
- GDP per capita (PPP): $12,400 (2017 est.)
- Inflation rate (cpi): 3.8% (2017 est.)

h) **Japan:**
- Country of Asia
- Capital: Tokyo
- Year of Independence: N/A
- Total area: 377,915 sq km
- Population: 125,507,472 (July 2020 est.)
- Language: Japanese
- Literacy: 99%
- Unemployment rate: 2.9% (2017 est.)
- Government: parliamentary constitutional monarchy
- GDP (PPP): $5.443 trillion (2017 est.)
- GDP (real growth rate): 1.7% (2017 est.)
- GDP per capita (PPP): $42,900 (2017 est.)
- Inflation rate (cpi): 0.5% (2017 est.)

i) **South Korea:**
- Country of Asia
- Capital: Seoul
- Year of Independence: 1945 from Japan
- Total area: 99,720 sq km
- Population: 51,835,110 (July 2020 est.)
- Language: Korean, English
- Literacy: 97.9%

- Unemployment rate: 3.7% (2017 est.)
- Government: presidential republic
- GDP (PPP): $2.035 trillion (2017 est.)
- GDP (real growth rate): 3.1% (2017 est.)
- GDP per capita (PPP): $39,500 (2017 est.)
- Inflation rate (cpi): 1.9% (2017 est.)

**j) South Africa:**
- Country of Africa
- Capital: Pretoria
- Year of Independence: 1910
- Total area: 1,219,090 sq km
- Population: 56,463,617 (July 2020 est.)
- Language: Zulu (official) 24.7%, isiXhosa (official) 15.6%, Afrikaans (official) 12.1%, Sepedi (official) 9.8%, Setswana (official) 8.9%, English (official) 8.4%, Sesotho (official) 8%, Xitsonga (official) 4%, siSwati (official) 2.6%, Tshivenda (official) 2.5%, isiNdebele (official) 1.6%, other (includes Khoi, Nama, and San languages) 1.9% (2017 est.)
- Literacy: 87%
- Unemployment rate: 27.5% (2017 est.)
- Government: parliamentary republic
- GDP (PPP): $767.2 billion (2017 est.)
- GDP (real growth rate): 1.3% (2017 est.)
- GDP per capita (PPP): $13,600 (2017 est.)
- Inflation rate (cpi): 5.3% (2017 est.)

**k) United States of America:**
- Country of North America
- Capital: Washington, DC
- Year of Independence: 1776 from the United Kingdom
- Total area: 9,833,517 sq km
- Population: 332,639,102 (July 2020 est.)
- Language: English only 78.2%, Spanish 13.4%, Chinese 1.1%, other 7.3% (2017 est.)
- Literacy: 99%
- Unemployment rate: 4.4% (2017 est.)
- Government: constitutional federal republic
- GDP (PPP): $19.49 trillion (2017 est.)
- GDP (real growth rate): 2.2% (2017 est.)

- GDP per capita (PPP): $59,800 (2017 est.)
- Inflation rate (cpi): 2.1% (2017 est.)

There are two-line charts on the following page (Line chart 1 and Line chart 2). Line chart 1 graphically presents an economic growth of world and above mentioned 11 countries. The size of real GDP is on Y axis, year is on X axis. Line chart 2 is a graphical representation of World's real GDP and sum of real GDP of the 11 countries from G-20. It provides an easy comparative picture.

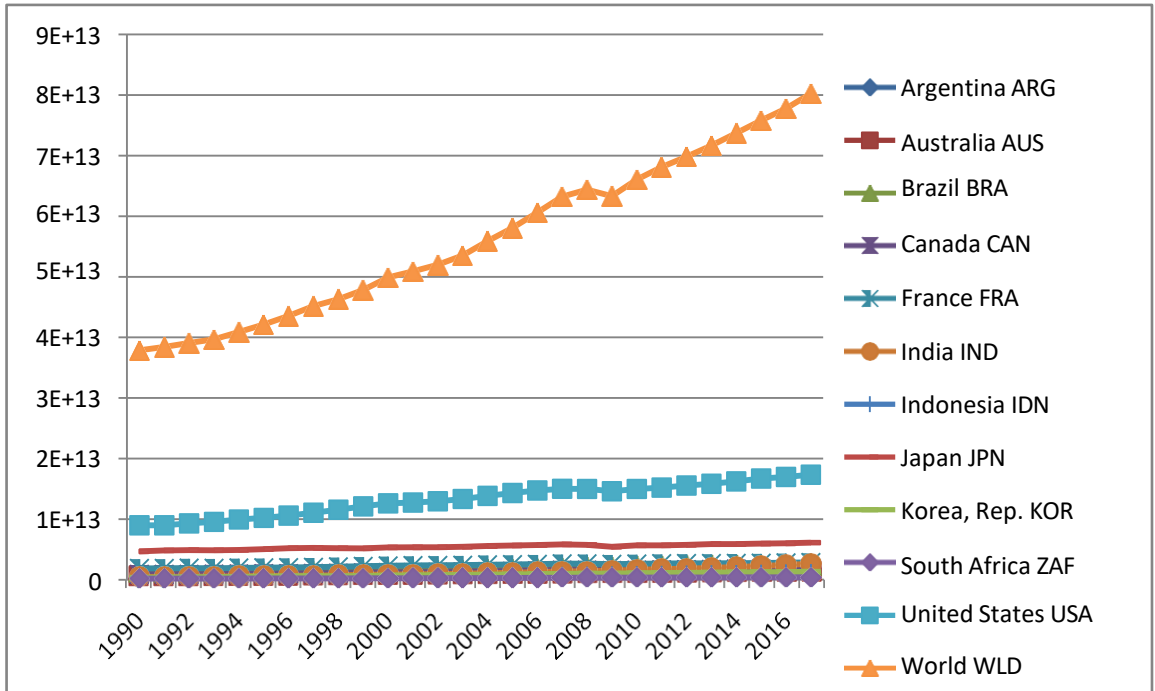

**Line chart 1**

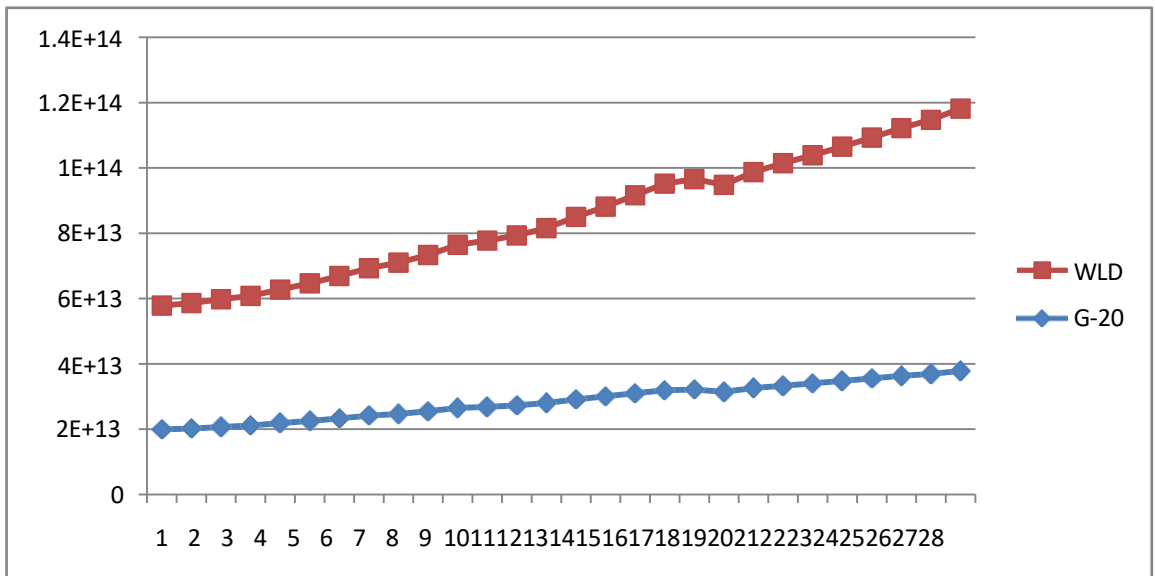

**Line chart 2**

## 1.3 Objectives

I have tried to examine the effects of trade openness and tariffs on trade on economic growth of the 11 of G-20 countries. Those 11 countries have been chosen purely based on data availability. This study tries to examine general impacts of trade openness and trade tariffs on economic growth. An important part of this study is to see the relationship and linkages between trade and economic growth for examined countries. As gross domestic product (GDP) consists of other important variables, empirical study of those is also an objective.

Most of the literature on relationships between trade openness and growth has left impact of direct tariffs aside. This is an attempt to study whether it has significance on economic growth and causality with the dependent variable. This topic is much debated and studied continuously. But the resolve is far from being reached. Different studies found different significant results. Individual country's time series analysis and panel data on group of countries have been carried out in the past. As the global recession, following global Covid-19 outbreak, has already started taking its tolls, countries will investigate every available tool in its disposal to cushion the shock, allocate the scarce resources and foster growth in long run. Another objective of this study is to suggest policy recommendations for economic growth by finding effective variables and causations.

I check causality directions and cointegrations of main variables, growth, and trade openness. Robustness of trade and tariff as important determinants of GDP growth will be examined in this case study.

### 1.3.1 Summary Objectives

A summary of main objectives is as follow:
- Examine the impact of trade openness on those 11 countries' economies.
- Examine the impact of tariffs on economic growth.
- Examine the relationships among trade, tariffs, and growth for examined countries.
- Examine the impacts of other variables, if any.
- Examine the real factors of growth and make appropriate suggestions.
- Propose policy suggestions to maintain economic growth amidst crisis, e.g. global recession.

## 1.4 Research Questions

This study investigates a few questions related primarily to connections among trade, tariffs, and economic growth. Some questions also investigate empirically the said relationship. Questions are:

- Is there a positive relationship between trade openness and economic growth of those 11 countries of G-20 group?
- What is the impact of trade tariffs on trade growth and economic growth?
- What is the causality direction of trade and economic growth?
- What is the nature of the impact trade openness having on economic growth of countries?
- What are the impacts of other variables along with trade and tariffs have on economic growth?

## 1.5 Contribution

I intend to empirically try and test linkages among trade openness (as percentage to GDP), economic growth and tariffs on trade (as percentage to total revenue). Although the relationship between growth and trade has been researched a lot, the impact of tariffs on trade and growth is not much studied. My expectation is that this study will make some important policy suggestions not just for the examined countries, but for most of the countries. Those may contribute to times of crisis like global recession and post-recession recovery, also for achieving right growth. This study shall lead to further empirical studies in this field to find in depth and clearer relationship among the variables.

## 1.6 Limitations

This thesis work will use secondary sourced data in its entirety. Limitations of this study range from sourcing of data to technical knowledge in statistics software as well as econometric technics. Reliable internet connection, modern software and logistics related to econometric technics were not always readily available.

## Chapter II Literature Review

### 2.1 Introduction

Wealth, economy, development, money, trade, and many other economic aspects have been intertwined with the evolution of humankind since the very early days. Mankind fought wars, struck deals, and established trade and commerce to harness benefits from those economic activities. These have not always been a win-win situation. Some nations got rich, some did not. A good number of economists and authors examined the causes behind this phenomenon. There are hundreds of books and several established theories on economic development, trade, commerce, and other related fields.

## 2.2 Theoretical Literature Review

Perhaps it would not be surprising if I mention economist Adam Smith (1723-90) and his "The Wealth of the Nations" in the first paragraph of literature review. Classical economists like Adam Smith who saw benefits in unhindered and barrier free trade as he argued that trade growth increases specialization in his famous literature work. This specialization leads to economies of scale and greater market for small domestic economies.

Work of Adam Smith mainstreamed trade as enhancer of welfare and growth which is still relevant, with much improvement from subsequent works. He saw trade as an effective mean to vent surplus home production. That vent is a widening of market which is overseas. By doing so, it also improves the division of labor and the productivity at home. However, critics point out to the impact this theory had on European colonialism and exploitation of colonies. But Lewis Hill and Betsy J. Clary concluded in their work that Adam Smith advocated for anticolonialism and anti-protectionism policies. He justified this through his literature works. Adam Smith was fundamentally opposed to the very idea of colonialism and barriers on international trade, which is a proponent of mercantilism.

Mercantilism was the dominant theory of economy and growth before being replaced gradually by Adam Smith's liberal market economy ideas. Mercantilist theories of International Trade see benefits in export surpluses, thus advocates for export over import and trade barriers. It argues that economic surplus is the fastest way to achieve economic development i.e. per capita income growth, sustainable development (Jozefina Semancikova, 2016). Therefore, tariffs and trade barriers are frequently imposed on imports. After the decolonization in post-WWII, newly independent countries took measures of import substitution policies and strategies to foster fastest growths. Those countries-imposed taxes on imported manufactured goods. For countries like the US, Germany, France, Soviet Union and China, protectionism brought positivity, though the UK was open to international trade in 18$^{th}$ and 19$^{th}$ centuries (Robert Baldwin, 2004). Newly independent countries replicated this model in the 60s and 70s by imposing tariffs, exchange rate mechanisms, etc. However, this trend did not follow through the following decades. In the late 1970s and 80s and into 90s, more and more developing countries shifted their policies from their previous inward-looking approach to outward looking approach (Robert Baldwin, 2004). It was achieved because of tremendous economic growths achieved my East Asian countries who had adopted that policy, and inability of inward-looking approach from delivering sustainable economic growth as shown in works of Michaely and Choksi (1991), et al. The policy model adopted is labelled, the "Washington consensus", that includes fiscal reform and discipline, stability of price, liberalization of trade and privatization of state enterprises. Both foreign and private investments and capital movements were unhindered. Many development economists supported the policy model which is based on open market economy and global financial and economic integration for growth and prosperity.

David Ricardo's Theory of Comparative Advantange refers to economy's ability to produce goods and services at a lower opportunity cost than that of trade partners. The Ricardian model suggests economies can benefit from specialization in sectors where they have comparative advantages. The Heckler-Ohlin model of international trade shows that specialization based on factor endowment can lead to static productivity and economic benefits.

Another approach which is one of the forefronts in economic growth theories is Neoclassical economics. It champions the idea of market determined by laws of supply and demand. Main assumptions are people are rational, individual, and firm seek utility and profit maximization, respectively and people act independently and informed. Based on this theory and assumptions, the Solow-Swan Model attempts to explain long term economic growth by examining capital accumulation, labor growth and diffusion of technology. The Exogenous Growth model stems from this neo-classical model as long run growth is impacted by external factors like technological progress (Sergio Rebelo, 1991). However, rates of saving and technological progress could not be explained. Endogenous growth model steps in here.

Endogenous growth model explains the economic growth rate by examining and giving importance to technology development, human capital, knowledge creation, and positive outcomes of knowledge-based economy. These importances are stressed out by Robert Baldwin (2004). He pointed out that growth depended on trade. He also mentioned the work of Grossman and Helpman (1991) on endogenous growth had introduced concepts such as knowledge spillovers resulting from trade, foreign direct investment as well as the replication of the products of foreign producers. Import protection generally reduces growth rates under these formulations, he added. Further, the new growth theory has provided important insights into an understanding of the relationship between trade and growth. These models exhibit international trade can promote economic growth through employment generation, income distribution, technology spill over and external stimulation.

Elahana Helpman (2004) commissioned a comprehensive approach to economic growth in her 2004 work. It was a blend of exogenous and endogenous effects of growth. She proposed that economic growth had 4 themes: 1) Technological and institutional factors affect the rate of accumulation of physical and human capital. Though can partially describe growth, the importance of these two cannot be offset. 2) Knowledge, its creation, research and development, institutional factors related to knowledge creation are important to understand total factor productivity. 3) Countries' growth rates are interdependent. Free flow of knowledge across national borders, FDI, international trade and capital flow affect the motives to innovate, to replicate, and to use new technologies. 4) Economic and Political institutions are important as these have direct influence over capabilities to adapt changes in international trade and economy.

**2.3 Empirical Literature review**

This topic has been contested heavily in the realm of trade, trade gains and economic growth. It is yet to be established, if not ambiguous. Different researchers used different variables. Some are commons, some are seldomly used. All those studies paint a complex and intertwined picture of economic growth and trade.

The first to address the issue were Scitovski, Scott and Balassa (1993). Their study has 2 sub-divisions: a) cross-country study of relations between trade and growth, b) investigation of experiences countries had gone through of trade reforms. Early studies infered conclusion that trade generally accelerated growth, whilst more recent studies casted doubt on these findings suggesting growth led trade.

15 | P a g e

Azam Chaudhry (2011) studied the relationship between trade and growth from institutional quality aspects. His model shows that countries with higher quality institutions gain more than countries with weaker institutions. When a government is threatened by international trade, it can raise tariffs, block off imports, which increases revenue and stability but reduces long term growth. However, it gives argument about tariff-growth paradox- countries experience high revenue and high growth at initial stage which fades off at later stages.

Anthony P. Thirlwall (2000) in his paper, titled "Trade, Trade Liberalization and Economic Growth: Theory and Evidence", suggests that gains from trade are not equally shared, i.g. in Customs Union. However, he finds that generalized trade liberalization in terms of tariffs reduction or withdrawal improves overall economic growth. His paper gives policy recommendation for export-led growth for developing countries, however, its effectiveness depends on characteristic of demands and production of goods.

Trade liberalization and openness can impact positively on economic growth rate, shown by Roman Wacziarg and Karen Horn Welch (2008). They examined data from 1950-98 and found that countries which liberalized their trade regimes enjoyed 1.5 percentage points higher than pre-liberalization growth rates. Post-liberalization investment rose by 1.5-2 percantages, conforming to past studies. Trade liberalization and openness benefitted economic growth, however, there were large differences among countries. Subsequent cross-country statistical studies found positive relations between export and economic growth. However, many differences among research in technics and issues examined have made it difficult to reach a conclusion. Srinivasan and Bhagwati in their 2001 study slammed cross-country regression analyses as the basis of deducing the relationships between trade openness and growth, pointing out weak models, poor data, and inappropriate technique (Robert Baldwin, 2004).

What makes long-term economic growth possible is studied by Gene M. Grossman and Elhanan Helpman in their 1994 paper on the theory of growth. They first examined neo-classical models, which advocate protections from externality, and theory of endogenous growth. They empirically found long run economic growth largely depended on endogenous technological progress. Dan Ben David and Micheal Leowy in 1998 paper argued from their long run perspective that unilateral and multilateral liberalization did generate a positive impact on steady growth of participating countries. Their model emphasized that knowledge gained from trade could have positive outcome on gains in income and growth rate in long run.

Integration of both trade openness, trade-gdp ratio, and effective imposed tariff rate as interest variables are conducted by Sabina Silajdzic and Eldin Mehic (2018) in their study of "Trade Openness and Economic Growth: Empirical Evidence from Transition Economies". Intriguing enough, they point out the ambiguousness of this proposition and divergent results across countries. Real GDP per capita is a dependent variable and effectively applied tariff rate and trade share to GDP are two interest variables among others. Results show that trade openness measured by trade share to GDP indicator is positively significant but very small value, whereas tariff has negative coefficient though insignificant.

Trade openness, along with labor participation rate and high level of domestic investment, have positive significant impact on economic growth, Dang Van Dan and Vu Duc Binh (2018) found in their paper. Inclusion of labor is an interesting development in the study of growth. Quality of labor and labor participation rate are very much country specific and dependent on various aspects. Indeed, this thesis paper will not examine those dependencies. Authors of the above study found that inflation, money supply and interest rate had negative effects on growth, particularly developing countries.

China development model is a sought after and much appreciated model of development. This model has been adopted by various nations, with varying results. China model which is dominated by outward looking approach and economic growth benefitted from increased exports. Yanqing Jian (2014) found technology diffusion helped China tremendously. Trade openness promoted total factor productivity in China. This effect combined with technological spillovers through exposure to advanced economies. He also found regional trade growth had positive significant impact on regional growth.

Trade and trade restricitions, i.e., tariffs, have been here since ancient times, but their effects are yet to be conclusive. Nathan Nunn and Daniel Trefler (2010) studied the topic and showed in the study that "skill bias" of country's tariff regime has positive correlation with economic growth. They examined GDP per capita and found long term positive correlation with the said variable while controlling other variables such as region fixed effects, initial production structure, per capita income, human capital, and investment. The "skill bias" is ratio of skilled workers and unskilled workers in industries. They examined GDP per capita and found long term positive correlation with the variable. But its alternative explanations found extensive rent seeking activities in the economy. Tariffs to protect skill-based industries increase rent and growth through endogenous activity.

In the paper titled "The determinants of Foreign Direct Investment: A Panel Data Analysis for the emerging Asian Economies" ATM Faruq (2023) has examined institutions, political along with economic factors behind international trade. The study found Business Disclosure Index impact FDI significantly. That finding indicates foreign business entities careful about the regulations, legal protection, tax burden etc. while considering mobilizing fund into the emerging Asian countries.

I, particularly, want to shed light on the response and impacts of economy of the United States to the trade disputes and tariffs. Tax Foundation assessed the Trump administration's tariff regime which has been in forefront of global trade dispute with various countries, chiefly China. Its study found that total imposed tariff was US$79.96 bn as of February 2020. It costs long run GDP of -0.23% and wages -0.15%. Their key findings include a) imposed tariffs, along with retaliatory actions, will reduce economic growth and employment. b) threats of further imposition of tariffs will reduce GDP by 0.24% further and wages by 0.17%. Their study explicitly finds an inverse relation between raised tariffs and economic growth. Still there are ominous signs of further deterioration of the situation in the coming days. It is presumed that after the ongoing global pandemic of Covid-19 is over, protectionism shall rise and take global trade regimes back to pre-war status.

Janhavi Shanker Tripathi (2016) used time series analysis to examine trade and growth dynamics for G-20 countries. He took a countrywide study approach and found different results for different countries. He used the natural log value of GDP as dependent variable and trade-gdp ratio as independent variable. The result is varying. He found that some countries have causal linkages from trade openness to economic growth, but very few in other way around. He pointed out to the other factors which were out of the study consideration such as geographical settings and distances, cultural impacts, governance, varying economic policies.

Qazi Adnan Hye, Shahida Wizarat and Wee-Yeap Lau (2013) examined trade-led growth hypothesis for South Asia with 4 (four) estimations: 1) export-led growth, 2) import-led growth, 3) growth-led import, and 4) foreign trade deficits. They employed ARDL approach to examine.

long run relationship among export, import and economic growth. Granger causality test was used to find out direction of causality. Results were mixed with all, but one country tested positive to export led growth while all countries have positive relation with import led growth. Results showed that domestic and foreign markets demand contribute to growth and employment. It suggested to cater domestic demand in times of recession.

Md. Abu Hasan, Md. Sanaullah, Mir Khaled Iqbal Chowdhury, and Anita Zaman in their 2017 paper examined trade-led growth and growth-led trade hypothesis in Bangladesh by investigating real GDP and real export-imports. To test short and long run relationships, ARDL error correction model was used. In their study, they found empirical evidence that trade liberalization significantly benefited economic growth of Bangladesh.

Bishnu Kumar Adhikary (2011) found interesting result in his "FDI, Trade Openness, Capital Formation and Economic Growth in Bangladesh: A linkage Analysis". His time series analysis on the said subject found long term positive correlations among growth, FDI and capital formation, whereas openness had negative but diminishing impact. He gave findings of Grossman & Helpman (1991), and Barro & Sala-I-Martin (1995) that higher degree of openness creates greater capability to absorb new technologies, and this capability leads to grow more rapidly. Less outward oriented countries lack this capability. In contrast, he gave Edwards' (1998) argument that the rate of growth in the poorer countries was not solely dependent on openness, but rather on their basic and primary knowledge. He, however, pointed out to the fact the inconclusiveness of link between trade openness and economic growth. His dependent variable was natural log value of real GDP of Bangladesh. Natural logarithm of trade over GDP, FDI and capital formation were independent and indicating variables.

I encountered a primary obstacle with empirical studies of international trade and growth link is how to measure trade openness or trade growth. The most used approach is to use the total trade volume, which is exports plus imports, and GDP ratio. Many empirical studies estimate positive effects on growth from trade liberalization and openness. The size of these effects, however, is often very small.

Mohamed Fenira (2015) in his empirical literature studied several trade openness indicators. He argued that even if those indices are designed to measure the same relationship of trade and growth, there were divergent results about the impact of trade has on growth, similarly tariff has on growth. The divergence came from different apprehension of trade openness and the manner of conceptualization. In his paper, he demonstrated that trade liberalization has a weak contribution in economic growth. Variable of interest was trade-gdp ratio (Export+Import/GDP).

Mr. Fenira also investigated interesting empirical work of Warner (2003). Warner in his work argued that weighted middle tariff rates had negative impacts on the growth, capital formation and intermediate goods. To its critic, Rodriguez in 2006 failed to replicate the result of Warner and concluded that trade policy was not a very strong indicator of growth. Here, it would be appropriate to mention that Warner and Rodriguez both worked with tariff rate as interest variable and found negatively correlated with growth, though not significant.

Yanikkaya (2002) proposed 5 different categories of trade openness: 1) Share of trade to gross.

domestic products (GDP), 2) Average tariff rates, export taxes, total taxes on trade, 3) Exchange rate, 4) Trade regimes and arrangements, and 5) indices of trade orientations in his work on trade openness and growth. I have used 2 measures, trade share to GDP and weighted tariff percentage on imports of all products as my interest variables in my study.

Ghoshal (2015) tried to investigate the causal relationship between trade and growth in India in her paper. She put special emphasis on the impact of introduction of various trade regimes and agreements. She examined annual GDP and total export of India to fathom the impacts.

Effective tariff applied on imports is regarded as a direct barrier. Some researchers have used it as an indicator of trade openness, as discussed above, and some researchers have used this as trade closeness indicator. David N DeJong and Marla Ripol (2006) examined tariff rates as a direct trade barrier measure. Their findings showed two important evidences: marginal effect of tariffs on trade was declining and (2) negative relationship between tariff and growth could only be seen among the rich countries.

A conclusion can be made at this stage after going thrugh various brief descriptions of literature of past works on trade-growth relations. It is more that apperant that most studies supported the positive relationship of trade openness and growth notion. However, it should not offset the econometric difficulties present in cross country analysis. These challenges arise from measurement problems, variables which possibly are endogenous and omitted variable bias. These difficulties generally dwell in trade openness measures. Although most studies find a positive relationship on average, they also stress the existence of heterogeneity in the effect of outward orientation on growth. Kyrre Stensnes (2006) mentioned recommendation of Rodrik and Rodriguez (2001) on this issue. Rodrik and Rodriguez suugested to investigate contingent relationship of trade openness and growth, which is basically country specific nature of trade policy.

**Chapter III Methodology**

**3 Methodology**

The above examination of various studies points out to the inconsistency observed in the empirical results examining the correlations among trade openness, trade barrier and economic growth. This may have an explanation that possible methodological deficiencies and complications in developing a perfect empirical model to examine the impact of trade openness and tariff on economic growth lead to inconsistency in results. As shown in some researchers' work that trade openness has an endogenous and dynamic nature which needs to be integrated into the empirical model. I have tried to minimize or remove such effects of endogeneity and other unobserved factors by employing econometric techniques and the best of my knowledge. I have discussed in detail in this matter in section number 3.4. The examined countries are from G-20, they share broad similarities in economic sizes and economic patterns. Time and country effect have also been considered in model.

### 3.1: Econometric Models

Investigation of relationships among trade openness, tariffs and economic growth has been done by a linear regression model. The model is:

$$Y(ln\_rgdp)_{ct} = \beta_0 + \beta_1 (trgrowth)_{ct} + \beta_2(tariff)_{ct} + \beta_3(ln\_govtexp)_{ct} + \beta_4(fdi)_{ct} + \beta_5 (domcap)_{ct} + \beta_6 (hh)_{ct} + \beta_6 (ln\_tl)_{ct} + \varepsilon_{ct} \quad\quad (1)$$

Whereas:
  a) $\beta_0$ = Intercept constant.
  b) ln_rgdp = Natural log value of real GDP in 2010 US$[1] which is inflation adjusted.
  c) trgrowth = Trade percentage of GDP (Sum of export and import relative to GDP), indicator of trade openness.
  d) tariff = Average of effectively applied tariff rates weighted by the product import shares. An indicator of trade openness as well as a direct barrier.
  e) fdi = Foreign Direct Investment, net inflows (percentage of GDP).
  f) domcap = Domestic Capital formulation (percentage of GDP).
  g) ln_govtexp = Natural log value of total government final consumption expenditure. Rational behind taking natural log of final government expenditure is getting the change effects on real GDP.
  h) hh = Household final expenditure percentage (%) to GDP.
  i) ln_tl = Natural log value of total labor, aged 15 and older.
  j) $\varepsilon_{ct}$ = Error terms, for unobserved data.

**Dependent variable:** ln_rgdp

**Independent variables**: trgrowth & tariff (interest variables), ln_govtexp, fdi, domcap, hh & ln_tl are control variables.

### 3.2 Data and Variables

Data selection is overwhelmingly important in a regression analysis. I explain and justify the choices I made in compiling a dataset in this section.

  I. **Data Source**

  All the relevant data on those countries have been obtained from secondary sources. World Development Indicators of The World Bank Group is the main source of all but one data. UNCTAD and WTO online libraries have been browsed for data cross check.

  II. **Data collection**

  Data has been collected through internet browsing. Data sources are free of charge. Compilation and cleaning up of data have been carried out offline.

---
[1] The use of the log value model allows to determine the responsiveness of economic growth to changes in the controlled variables used in the study.

### III.    Variables

a) Real GDP (2010 US$) or real gross domestic product (GDP) is a macroeconomic measure. It is the price adjusted value of economic output. I obtained data of GDPs of 11 countries from the World Bank world development indicators website. Countries' GDPs are constant 2010 US dollar. Natural log values have been taken in regression analysis to show changes in real GDP. As real GDP is inflation adjusted, inflation was not considered as control variable.

b) Trade openness, named in econometric model as growth, has been debated much by researchers based on its interpretation. I have considered and measured trade share to GDP which is the sum of export and import over GDP (X+M/GDP). I have chosen as it has been shown by other researchers that this index is not exposed to inconsistencies. Indeed, there may be some endogeneity in countries' trade share data which has been taken of in regression model. All the 11 countries is categorized on the basis of one criteria, and that is the sum of export and import over GDP. This variable denotes the trade share of GDP of a country c in the period of t.

c) Tariff variable represents Weighted mean applied tariff that is the average of effectively applied rates weighted by the product import shares corresponding to each country (World Bank WDI). Import weights were calculated using the United Nations Statistics Division's Commodity Trade database. Rationale behind using the traiff variable as indicator of openness as well as trade restriction can be found in works of some previous researchers like Edwards and Rodrik and Rodrigues, as described by Kyrre Stensnes. Edwards (1998) used tariff rates as direct measures of trade policy. Rodrik and Rodriguez advocated using tariff averages or coverage ratios for non-tariff barriers strongly. They gave arguement that these were the most direct measures of trade policy available. As it has been mentioned above that Warner and Rodriguez, et al worked with tariff rate as interest variable. The World Trade Organization has abolished quota system in international trade, largely, however. Rate of tariff can effectively showcase country's dependency on tariff generated by international trade. Therefore, I have decided to use tariff rate as variable in my regression.

d) FDI or foreign direct investment has been represented as a percentage of a country's GDP. This is a control variable. It has been used with other variables in a pooled dataset. Logic is previous studies have found correlation between growth and FDI.

e) Domcap is domestic capital formation in a country. Data is percentage to GDP. It is also a control variable like FDI.

f) Ln_govtexp is a natural logarithm of total government expenditure of a country. It is an important control variable. Government expenditure or consumption is part of a country's total consumption, an important part of GDP.

g) Hh is percentage of household final consumption expenditure in percentage to GDP. It along with government consumption form total consumption of a country.

h) Ln_tl in natural log value of total labor participants of a country. The long value of total labor can show changes in labor participation rate. It is also a control variable.

i) $\varepsilon_{ct}$ is an error term. An error term is included to capture and understand the variation in the dependent variable. Error term is independent to independent variables as it shows the changes in dependent variable which cannot be explained by interest and control variables. The structure of error term depends on whether the model in ordinary least square or OLS, fixed effect (FE) or random effect (RE) model. An appropriate test has been conducted to find the best suited.

### 3.3 Estimation Techniques

My study will use panel data set ranging from 1990 to 2017 of the 11 countries- Argentina, Australia, Brazil, Canada, France, India, Indonesia, Republic of Korea, Japan, South Africa, and the United States. Study will employ most appropriate regression models- Ordinary Least Squares and Fixed Effects or Random Effects whichever is appropriate. Stata software will be used to carry out the research work. It is a very able economic software on its own.

### 3.4 Dealing with Endogeneity

In my econometric analysis I tried to capture the problem of possible endogeneity of the trade openness variable. The problem of the possible endogeneity understates the OLS result and makes econometric model weak. This potential interdependence between trade variables and economic growth may lead to correlation between independent variables and error term. But one of OLS assumptions is error term is independent of other variables. All of these may produce results for parameters which are inconsistent and biased. Concerning exogeneity assumption of OLS, empirical literatures suggest using lagged values of the independent variables with the purpose of examination of causality among economic growth *ln_rgdp* and interest variables (*trgrowth*, *tariff*). Substituting *trgrowth*$_{ct}$ with *trgrowth*$_{ct-1}$ effectively eliminates concern that *trgrowth*$_{ct}$ is endogenous to economic growth (*ln_rgdp*)$_{ct}$. So is with *tariff*$_{ct-1}$ in place of *tariff*$_{ct}$. Concerning the said matter, I employ regression analysis by taking one period lag (t-1) of interest variables and all other control variables. The model is the following:

$$Y(ln\_rgdp)_{ct} = \beta_0 + \beta_1 (trgrowth)_{ct-1} + \beta_2 (tariff)_{ct-1} + \beta_3 (ln\_govtexp)_{ct-1} + \beta_4 (fdi)_{ct-1} + \beta_5 (domcap)_{ct-1} + \beta_6 (hh)_{ct-1} + \beta_6 (ln\_tl)_{ct-1} + \varepsilon_{ct} \dots\dots\dots (2)$$

In the second equation, *rgdp*$_{ct}$ represents the growth of economy of country c in period t. *trgrowth*$_{ct-1}$ is the trade growth in country c during the previous period, *tariff*$_{ct-1}$ is one-period lagged of *tariff*, similarly one-period lagged the other controls are taken. The main rationale to take one year lagged of explanatory and control variable is to control the reverse causality biasness which is one of the main causes of endogeneity.

# Chapter IV Preliminary Result

This chapter will show and discuss the estimation results found in regression analysis using econometric software.

## 4.1 Descriptive Statistics and Correlation coefficients

Table 1 represents summery statistics of the variables. As it can be seen in the table that highest number of observations (*n*) is 308. However, *trgrowth* and *tariff* have less than that as data was not available. The meaning of *ln_rgdp* is 27.893 which is average of all values. Mean of *trgrowth* is.
43.218. The *tariff* is 6.330. Standard Deviation (Std. Dev) shows how far observations are from the averages. Min shows the minimum value of the variable. Max represents the maximum value of the variable.

| Variable | Obs | Mean | Std. Dev | Min | Max |
|---|---|---|---|---|---|
| ln_rgdp | 308 | 27.89324 | 1.118896 | 26.04005 | 30.48453 |
| trgrowth | 308 | 43.21755 | 19.54014 | 13.75305 | 110.0001 |
| tariff | 278 | 6.330468 | 5.872893 | .89 | 56.36 |
| ln_govtexp | 305 | 26.10549 | 1.222347 | 24.07912 | 28.55136 |
| fdi | 308 | 1.701649 | 1.519133 | -3.618815 | 9.201639 |
| domcap | 308 | 24.28253 | 5.997038 | 10.85391 | 41.93083 |
| hh | 308 | 59.514 | 5.496156 | 48.09497 | 81.85653 |
| ln_tl | 308 | 17.59707 | 1.100201 | 15.95545 | 19.99807 |

Table 1: Descriptive statistics of variables

Table 2 presents data of correlation matrix of variables. My dependent variable *ln_rgdp* has a negetaive correlation with *trgrowth* which represents trade openness. *ln_rgdp* and *trgrowth* are negatively correlated with *tariff*. *Ln_govtexp* is positively correlated with my dependent variable *ln_rgdp* and the value is quite high. *Ln_govtexp* is negatively correlated with my interest variables. *Fdi* is negatively correlated with both *ln_rgdp* and *tariff* but positively with *trgrowth* and *ln_govtexp*. *Domcap* is only negatively correlated with *ln_govtexp* and *fdi*. *Hh* or household expenditure percentage is positively correlated with *tariff* and *fdi*. *Ln_tl* is only negatively correlated with *trgrowth* and *fdi*.

|  | ln_rgdp | trgrowth | tariff | ln_gov~p | fdi | domcap | hh | ln_tl |
|---|---|---|---|---|---|---|---|---|
| ln_rgdp | 1.0000 | | | | | | | |
| trgrowth | -0.3439 | 1.0000 | | | | | | |
| tariff | -0.3718 | -0.2466 | 1.0000 | | | | | |
| ln_govtexp | 0.9694 | -0.3345 | -0.4069 | 1.0000 | | | | |
| fdi | -0.0484 | 0.0582 | -0.1597 | 0.0017 | 1.0000 | | | |
| domcap | 0.0297 | 0.2525 | 0.1176 | -0.1076 | -0.1557 | 1.0000 | | |
| hh | -0.0313 | -0.4954 | 0.2196 | -0.0865 | 0.0532 | -0.5339 | 1.0000 | |
| ln_tl | 0.4662 | -0.3658 | 0.2640 | 0.2973 | -0.2065 | 0.3003 | 0.2929 | 1.0000 |

Table 2: correlation matrix

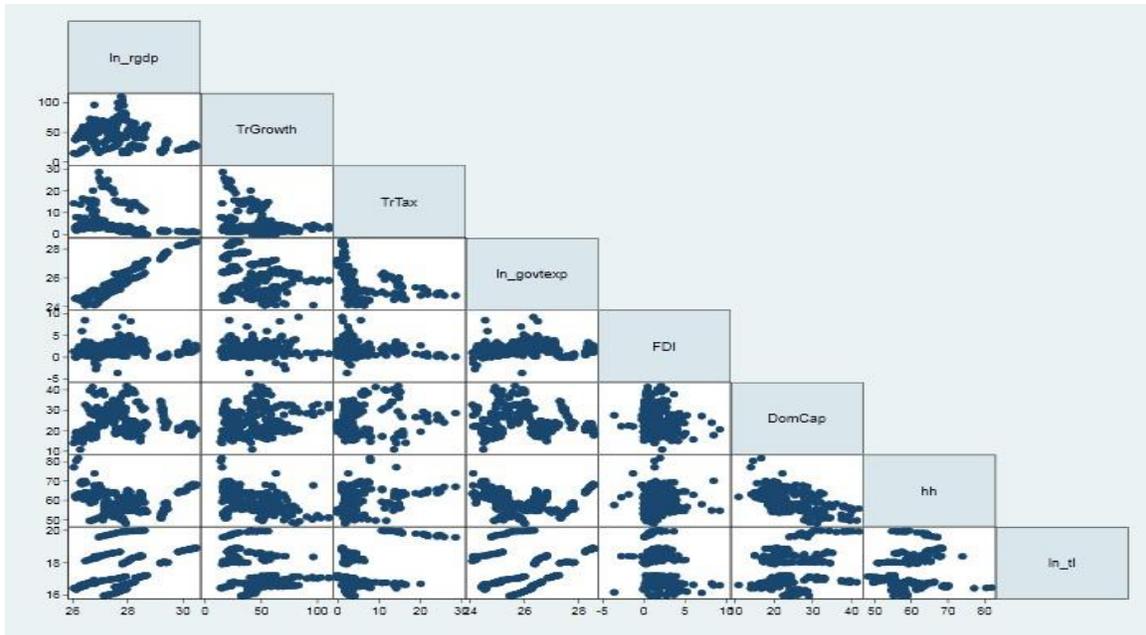
Graph 1: Correlation matrix

## 4.2 Regression Results

I will show results of econometric equation model (1) in Table 3. I have used Ordinary Least Squear or OLS, OLS robust, Random Effects, Fixed Effects and Fixed Effects robust. I have conducted the Hausman test to find whichever is more appropriate for estimation. I will discuss this issue later. The first 2 columns show estimation results of OLS and OLS robust regression. Robust method has been used to remove heteroskedasticity of variables. Values in parentheses are standard deviation. The number of observations is 277. Values of R-squared vary across regression methods.

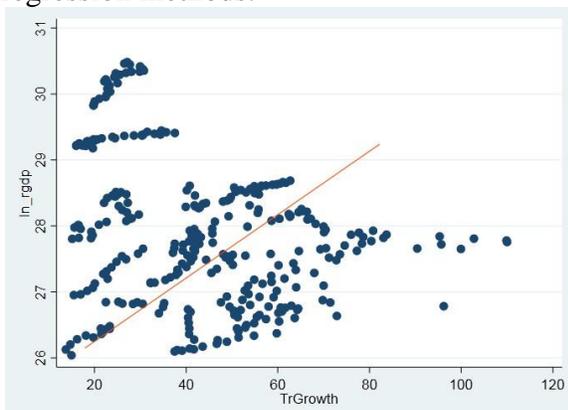
Graph 2: scatter plot of trade share and real GDP growth

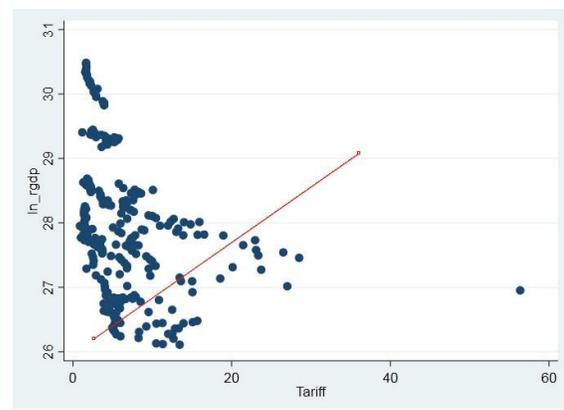
Graph 3: scatter plot of tariff and real GDP growth

|  | 1<br>(OLS) | 2<br>(OLS-robust) | 3<br>(Random Effects) | 4<br>(Fixed Effects) | 5<br>(FE-robust) |
| --- | --- | --- | --- | --- | --- |
| VARIABLES | ln_rgdp | ln_rgdp | ln_rgdp | ln_rgdp | ln_rgdp |
| trgrowth | 0.00135*** | 0.00135*** | 0.00204*** | 0.00186*** | 0.00186** |
|  | (0.000508) | (0.000484) | (0.000450) | (0.000424) | (0.000796) |
| tariff | -0.0121*** | -0.0121*** | -0.00382*** | -0.00223* | -0.00223 |
|  | (0.00163) | (0.00210) | (0.00121) | (0.00115) | (0.00226) |
| ln_govtexp | 0.859*** | 0.859*** | 0.829*** | 0.721*** | 0.721*** |
|  | (0.00941) | (0.0107) | (0.0195) | (0.0268) | (0.0875) |
| fdi | -0.0129*** | -0.0129*** | 0.00735** | 0.00548** | 0.00548* |
|  | (0.00479) | (0.00498) | (0.00287) | (0.00262) | (0.00273) |
| domcap | 0.0309*** | 0.0309*** | 0.00840*** | 0.00773*** | 0.00773** |
|  | (0.00203) | (0.00227) | (0.00158) | (0.00144) | (0.00334) |
| hh | 0.0276*** | 0.0276*** | 0.0102*** | 0.00938*** | 0.00938** |
|  | (0.00249) | (0.00203) | (0.00207) | (0.00190) | (0.00408) |
| ln_tl | 0.126*** | 0.126*** | 0.276*** | 0.619*** | 0.619*** |
|  | (0.0110) | (0.0123) | (0.0278) | (0.0553) | (0.178) |
| Constant | 0.919*** | 0.919*** | 0.525 | -2.619*** | -2.619 |
|  | (0.302) | (0.295) | (0.533) | (0.704) | (2.369) |
| Observations | 277 | 277 | 277 | 277 | 277 |
| R-squared | 0.989 | 0.989 |  | 0.958 | 0.958 |
| Number of countryname |  |  | 11 | 11 | 11 |

Table 3: regression results (Values in parentheses are standard daviation)

Estimation results of 1st column support the hypothesis that economic growth is positively correlated with interest variable *trgrowth* or trade to gdp share. It is significant at 1%. However, the coefficient of trade share to gdp is very low. It indicates that one unit increase in trade leads to 0.00135 unit increase in real GDP growth. Tariff has a negative correlation with economic growth as expected. Coefficient is significant at 1% statistics (p-value is less than 0.01). Negative 0.0121-unit change will occur in real GDP if 1 unit of tariff is increased. Government expenditure and total labor participation have quite high positive coefficients, 0.859 and 0.126 respectively. Both coefficients are significant at 1% level. This shows that a 1% increase in government expenditure will have a positive change of 0.859% in real GDP rate. Also, a 1% increase in labor participation rate will increase real GDP by 0.126%. The result of R-squared is quite high, 98.9% which means this model can capture the 98.9% variance in real GDP changes. The second column shows results of OLS robust testing of the model.

My pooled dataset in a panel dataset. Therefore, it is more appropriate to use either random effects model or fixed effects model. Let's investigate the third and fourth columns of table 3. Column 3 has estimation results of random effects model. Trade openness is positively correlated with economic growth. Its coefficient is 0.00204 which is significant at 1%. Tariff is negatively correlated; its coefficient is negative 0.00382 and significant at 1%. Column four (4) presents results of fixed effects regression. R-sqr value is 0.958 which means this model can capture 95.8% of variance in changes of real GDP. Co-efficient of intercept is negatively significant at 1%. My first interest variable trade openness or *trgrowth* is significant at 1%. Its coefficient is positive 0.00186 which means 1 unit increase in trade openness or share shall have 0.00186 unit increase of real GDP growth. The correlation is positive, although, value is low margin. Another interest variable *tariff* is negatively correlated with the real GDP growth, like OLS and random.

effects models. However, change can be seen at a significant level. Coefficient of *tariff* is significant at 10% statistics, unlike previous models. Negative coefficient 0.00223 indicates 1 unit increase in tariffs will decrease real GDP growth by 0.00223 unit. All the other control variables (government expenditure, foreign direct investment, domestic investment, household consumption and total labor participation) are positively significant at the 1% level. Observed weak relation between trade openness and economic growth has also been observed in previous studies carries out by Mohamed Fenira, 2015 and Janhavi Shanker Tripathi, 2016. Specially Mr. Tripathi pointed out to geographical and economic policies differences. On the other hand, low margin, yet negative coefficient of *tariff* variable has tendency to meet results from Sabina and Eldin, 2018. They had positive significant correlation between economic growth rate and tariff using least squears dummy variables (LSDV) in transition economies. However, that result changed in different model with same data. The synthesis of those studies implies that trade openness and tariff do not always have positive and negative impacts on economic growth as shown in previous studies. To carry further tests, one appropriate model has to be selected from random effects or fixed effects model.

**Hausman test**

To select the most appropriate model, I have conducted the Hausman test to find the best model. It tests whether error terms (Eit) are correlated with the regressors. Result of the test has been shown in the following Figure 1:

```
                ---- Coefficients ----
                 (b)          (B)         (b-B)        sqrt(diag(V_b-V_B))
                  fe           re        Difference           S.E.

    trgrowth    .0018643     .0020433    -.000179           .0001369
      tariff   -.0022329    -.0038178     .0015849           .0004058
   ln_govtexp   .721303      .8289506    -.1076476           .0225094
         fdi   .0054816     .0073499    -.0018683           .0004694
      domcap   .0077283     .0084034    -.0006751           .0002455
          hh   .0093795     .0102008    -.0008213           .0004362
       ln_tl   .6186383     .2762854     .3423528           .0546672

                   b = consistent under Ho and Ha; obtained from xtreg
       B = inconsistent under Ha, efficient under Ho; obtained from xtreg

   Test:  Ho:  difference in coefficients not systematic

              chi2(7) = (b-B)'[(V_b-V_B)^(-1)](b-B)
                      =    55.62
            Prob>chi2 =    0.0000
```

Figure 1: Hausman test

In Figure 1, we have the result of the Hausman test. The P-value of the test is less than 0.05 that means null hypothesis, difference in variables not systematic, is rejected and fixed effects model should be used. Fixed effects (FE) explore the relationship between dependent and indicator variables within a certain entity. Each entity has endemic characteristics which affect the outcome of regression analysis. This factor needs to be controlled. This is the assumption of correlation between indicator variables and error term. Fixed effects model controls all the time invariant characteristics in entities, so the estimated coefficients will not be biased by omitted or unobserved time invariant characteristics within variables and entities (in this case countries). Time invariant characteristics can be cultural, religion, race, etc.

**Heteroskedasticity testing**

After the Hausman test, the fixed effects model is finalized. I have run a test to check for heteroskedasticity in my model. The test is Modified Wald test.

```
Modified Wald test for groupwise heteroskedasticity
in fixed effect regression model

H0: sigma(i)^2 = sigma^2 for all i

chi2 (11)  =    1619.41
Prob>chi2  =     0.0000
```

Figure 2: Heteroskedasticity test

The P-value of the test is near to 0.000 or essentially zero. It means that we can reject the null hypothesis and accept alternative which is heteroskedasticity in model. To remove this error, I have run fixed-effect robust estimation. Results of FE robust are in column 5 of table 3. This model has produced R-sqr value of 0.958 which means it can capture 95.8% of variance in real GDP growth, same as fixed-effects model. Changes can be seen in coefficients of interest and control variables. Trade openness is positively correlated with economic growth. It is significant at 5% level. Coefficient is 0.00186. Coefficient estimate value and sign are identical with fixed-effects model. Only change has occurred at a significant level. One important change can be seen in another interest variable Tariff. It is still negatively correlated with growth, but not significant at 10% stat anymore. No change in Government Expenditure indicator variable estimate from fixed-effects. Foreign direct investment or FDI is now significant at 10% starting from the previous 5%, and positively correlated. Domestic capital formation or *domcap* has also seen a decrease in significance level from 1% to 5%, though retaining the same coefficient value and sign. Household expenditure has a positive coefficient value 0.00938 which is significant at 5% stat. Total labor participation remains same as fixed effects model, marking its important role in real GDP growth. Intercept coefficient is no longer significant at any stage. For the ease of readers, I have inserted table 3 in formatted version in table 4 with results of fixed-effects and robust estimates.

| VARIABLES | 4<br>(Fixed Effects)<br>ln_rgdp | 5<br>(FE-robust)<br>ln_rgdp |
|---|---|---|
|  |  |  |
| trgrowth | 0.00186*** | 0.00186** |
|  | (0.000424) | (0.000796) |
| tariff | -0.00223* | -0.00223 |
|  | (0.00115) | (0.00226) |
| ln_govtexp | 0.721*** | 0.721*** |
|  | (0.0268) | (0.0875) |
| fdi | 0.00548** | 0.00548* |
|  | (0.00262) | (0.00273) |
| domcap | 0.00773*** | 0.00773** |

|  |  | (0.00144) | (0.00334) |
|---|---|---|---|
| hh |  | 0.00938*** | 0.00938** |
|  |  | (0.00190) | (0.00408) |
| ln_tl |  | 0.619*** | 0.619*** |
|  |  | (0.0553) | (0.178) |
| Constant |  | -2.619*** | -2.619 |
|  |  | (0.704) | (2.369) |
|  |  |  |  |
| Observations |  | 277 | 277 |
| R-squared |  | 0.958 | 0.958 |
| Number of countryname |  | 11 | 11 |

Table 4: Fixed-effects and FE robust

Robustness of the model deals with heteroskedasticity problem on fixed-effect model. There is another issue to be tested and that is cross sectional dependence.

**Cross sectional dependence test**

Cross sectional dependence is an issue in panel data set with long time series or large number of years. To check cross-sectional dependence, Pesaran cross-sectional dependence test has been carried out. This test is used to examine whether the residuals are correlated across entities (Daniel Hoechle, Statacorp). Test result has been showcased in figure 3.

```
Pesaran's test of cross sectional independence =    -1.444, Pr = 0.1488

Average absolute value of the off-diagonal elements =      0.408
```

Figure 3: Pesaran cross-sectional dependence test

The test result shows the P-value is 0.1488 which is much higher than 0.05. Therefore, null hypothesis cannot be rejected, and there is no cross-sectional dependence.

Here, I want to mention about time fixed effect. I ran a test to see whether dummies for all year are equal to 0 (zero). If there are equal to 0, time fixed effect is not needed. Test result has P-value more than 0.05, that means null hypothesis cannot be rejected. Therefore, the time fixed effect was not run. Result of *testparm* is in annexure for the reference.

### 4.3   Endogeneity check

It was shown in both theoretical and empirical works that international trade-gdp ratio as trade openness has endogenous characteristics. Endogenous entity or variable can undermine test result and render econometric model less trustworthy. Therefore, this endogeneity in trade openness variable needs to be addressed.

As it was mentioned earlier that variables have been converted into lagged (t-1) values in econometric equation model (equation no. 2). By doing so and using econometric technique in Stata, it was possible to capture variables' impact on next year's growth estimation. Simply, previous year's data of all the independent variables were regressed to get present years.

economic growth rate.

| VARIABLES | 1<br>(OLS)<br>ln_rgdp | 2<br>(Random Effect)<br>ln_rgdp | 3<br>(Fixed Effect)<br>ln_rgdp | 4<br>(FE-robust)<br>ln_rgdp |
|---|---|---|---|---|
| L.trgrowth | 0.00139*** | 0.00167*** | 0.00157*** | 0.00157* |
|  | (0.000525) | (0.000466) | (0.000443) | (0.000724) |
| L.tariff | -0.0125*** | -0.00593*** | -0.00448*** | -0.00448** |
|  | (0.00169) | (0.00124) | (0.00118) | (0.00179) |
| L.ln_govtexp | 0.848*** | 0.790*** | 0.665*** | 0.665*** |
|  | (0.00982) | (0.0201) | (0.0286) | (0.0790) |
| L.fdi | -0.0137*** | 0.00587** | 0.00404 | 0.00404 |
|  | (0.00491) | (0.00293) | (0.00268) | (0.00265) |
| L.domcap | 0.0306*** | 0.00582*** | 0.00505*** | 0.00505 |
|  | (0.00210) | (0.00164) | (0.00149) | (0.00338) |
| L.hh | 0.0267*** | 0.00595*** | 0.00482** | 0.00482 |
|  | (0.00259) | (0.00213) | (0.00196) | (0.00387) |
| L.ln_tl | 0.135*** | 0.288*** | 0.649*** | 0.649*** |
|  | (0.0115) | (0.0282) | (0.0589) | (0.156) |
| Constant | 1.135*** | 1.700*** | -1.289* | -1.289 |
|  | (0.314) | (0.552) | (0.754) | (2.263) |
| Observations | 266 | 266 | 266 | 266 |
| R-squared | 0.988 |  | 0.954 | 0.954 |
| Number of countryname |  | 11 | 11 | 11 |

Table 5: regression results (Values in parentheses are standard daviation)

In table 5, regression results are shown in 1-4 columns. Each column shows estimations of individual regression methods. Column 1 has results of OLS regression. R-sqr value is 0.988 which means it can describe 98.8% of changes in outcome variable. Intercept is positive and significant at 1%. The coefficient of trade openness variable positively correlated and significant at 1%. Value of the coefficient does very nominally change from the value of coefficient in general OLS regression. Coefficient of tariff variable is also almost identical; change is very small. The sign of the variable does not change. In random effects regression results, trade openness coefficient is positively correlated and significant at 1%. However, coefficient value has come down to 0.00167 from 0.00186, in column 3, table 3. Coefficient value 0.00167 denotes that a 1% increase in present year's trade share to GDP shall increase next year's real GDP growth by 0.00186 percentage point, giving that all the other factors remain constant. Tariff variables are negatively correlated with economic growth. Coefficient value is 0.00593 and significant at 1%. One remarkable change occured in significance percentage of intercept coefficient. It is significant at 1% up from not significant at 10% in *t* random effects model. Column 4 has regression results of fixed-effects model. R-sqr value of this model is 0.954 which means it can describe 95.4% of changes in dependent variable real GDP growth. The trade openness variable is positively correlated and significant at 1%. The value of the coefficient is 0.00157 which denotes that 1% increase in present year's trade share to GDP will increase next year's real GDP growth by 0.00157 percentage point, giving that all the other factors remain constant (*ceteris paribus*). Coefficient value has come down from 0.00204 in *t* fixed effects model. Another interest variable, *tariff* is significant at 1% and has negative coefficient of 0.00448. Therefore, it can be said that 1%.

increase in present year's tariff rate will decrease next year's real GDP growth by 0.00448 percentage point. Foreign direct investment or FDI is not significant in lagged (t-1) fixed effect model, unlike previous results of equation 1. The coefficient of government expenditure has come down to 0.665 from 0.829 of fixed effects model from table 4. No changes in significance level and sign, however. The intercept coefficient is now negative and significant at 10%. All the other variables have seen little or no changes in values of coefficients, signs, and significant percentage. These findings are in parallel to the result of table 3 with the same sign and almost similar level of significance.

I have compared results of lagged (t-1) random and fixed effects models with the results from random and fixed effects models (equation 1). It is needed to select one suitable model from these two at this stage of test. Therefore, I have run a Hausman test in order to select like I did previously.

**Hausman test**

|  | (b) fe | (B) re | (b-B) Difference | sqrt(diag(V_b-V_B)) S.E. |
|---|---|---|---|---|
| trgrowth L1. | .0015708 | .0016705 | -.0000997 | .000157 |
| tariff L1. | -.0044808 | -.0059298 | .0014489 | .0004361 |
| ln_govtexp L1. | .664992 | .7903702 | -.1253782 | .0245037 |
| fdi L1. | .0040418 | .0058691 | -.0018273 | .0005109 |
| domcap L1. | .0050521 | .0058225 | -.0007704 | .0002635 |
| hh L1. | .0048201 | .005953 | -.0011329 | .0004483 |
| ln_tl L1. | .6491959 | .2882047 | .3609912 | .0589456 |

```
                b = consistent under Ho and Ha; obtained from xtreg
 B = inconsistent under Ha, efficient under Ho; obtained from xtreg

    Test:  Ho:  difference in coefficients not systematic

              chi2(7) = (b-B)'[(V_b-V_B)^(-1)](b-B)
                      =   54.21
            Prob>chi2 =    0.0000
```

Figure 4: Hausman test

The Hausman test result shows that P-value is close to 0.00, thus we can reject null hypothesis and select fixed effects (FE) model for further tests and analyses.

**Heteroskedasticity test**

After finalizing the fixed effects model for further analyses and study, Modified Wald test has been run to check for heteroskedasticity.

```
Modified Wald test for groupwise heteroskedasticity
in fixed effect regression model

H0: sigma(i)^2 = sigma^2 for all i

chi2 (11)  =    42297.43
Prob>chi2  =      0.0000
```

Figure 5: Heteroskedasticity test

Result shows P-value is near to 0.000 or essentially zero. It means that we can reject the null hypothesis and accept alternative which is heteroskedasticity in model. To remove this error, fixed-effect robust regression has been done. Estimations of this regression are in column 5 of table 5. R-sqr value is identical. Significance of trade growth variable over dependent variable is 10%, from 1% in t-1 FE model. Coefficient value is the same. Tariff has also lost its previous 1% significance and has new significance at 5%. Coefficient value remains same as t-1 FE model. The sign of coefficient is negative. The coefficient of government expenditure remains the largest among regressors. Lagged FE robust model produces same result as lagged FE model. Both models have very minimal data variance from FE and FE robust models (heteroskedasticity adjusted) of equation 1. Domestic investment is no longer significant at 1% level in lagged (t-1) fixed-effects robust model, even though, sign and coefficient remain same. Household expenditure sees the same fate as domestic investment. It is no longer significant at 5%. However, the total labor participation rate remains the same. It is positively correlated with the outcome variable economic growth. Its coefficient is 0.649 and significant at 1%. 1 percentage increase in labor participation rate from previous year will have 0.649% increase in economic growth rate of present year, giving that all the other factors remain constant (*ceteris paribus*). Intercept is negative and not significant at 10% level. Results of lagged fixed-effects robust model shows robustness of data and model.

**Cross sectional independence test**

Cross sectional dependence is an issue in panel data set with long time series or large number of years. To check cross-sectional dependence, Pesaran cross-sectional dependence test has been carried out. This test is used to examine whether the residuals are correlated across entities (Daniel Hoechle, Statacorp.com). The test result has been showcased in figure 6.

```
Pesaran's test of cross sectional independence =     0.694, Pr = 0.4877

Average absolute value of the off-diagonal elements =    0.377
```

Figure 6: Cross-sectional independence test

The test result shows the P-value is 0.4877 which is much higher than 0.05. Therefore, we cannot reject null hypothesis and there is no cross-sectional dependence.

## Chapter V

**Conclusion**

I have investigated the relations among trade openness, tariff rate and economic growth in this paper. Conventional wisdom is that trade openness leads to economic growth and income generation, and tariff cause peril on economic growth. I have discussed relevant literature and methodology of the test in chapter II and III. Existing literature on trade, tariff and economic relationships is ambiguous and not clearly established. That required the relationships to be tested empirically. Empirical results of it have been shown in chapter IV. Results have been shown in 2 segments of 2 equations. In the 1st segments, results have found a positive relationship between trade openness and economic growth of 11 (eleven) examined countries. However, even though results are significant and robust, low coefficient value indicates weak effectiveness of trade openness on economic growth. Whereas government expenditure and labor participation rate have strong positive correlations. Trade openness's weak link to economic growth indicates to the fact that increased outward orientations come with risks of external shocks, decreased domestic production, technological backwardness which can effectively bar countries from reaping the benefits of trade. It is to be noted that 5 (five) of the 11 countries examined are not in high income group. They are from the developing world, even though the economy size is big. Developing countries usually are laggard in technological innovation and adaptation, weak in institutions which are related to international trade and economy. As it has been shown in Kyrre Stensnes, 2006 paper that strong institutions of conflict management can make countries sufficiently prepared for external shocks and benefits from trade openness. Weak institutions fail to respond to changes and cause long-term losses. Therefore, spillover effects from trade and gains from integration to international trade largely depend on factors like efficiency, institutions, human capitals, and exchange rate, etc. Overall, it can be said that trade openness and economic growth relationships are not a singular and simple one, rather multifaceted nature of the relationships largely depend on differences in the levels of institutional development, size of the economy and technological proficiency. This finding has important theoretical and policy implications.

The above result cannot and should be interpreted into a blanket universal suggestion for rapid trade liberalization and opening borders as a fast means for economic growth. Rather country specific characteristics and heterogeneity of trade liberalization should be considered. Countries with strong institutions may open faster. The results have also shown importance of other macroeconomic factors such as government expenditure, labor participation and quality, domestic investment, and household expenditure as main drivers of economic growth. Relevant policies should be viewed and formulated before formalizing priorities. Future country specific studies can be conducted to appraise those factors' effects, for example, government expenditure and inflation, education rate and labor quality.

Results have shown that tariffs, another indicator of trade openness and a barrier, has negative correlation with economic growth. As like of trade openness, its low value coefficient though highly significant show weak influence over economic growth. This weak influence is particularly strong in same year tariff-growth relations. Results are at par with many empirical literatures. However, this is not the proposal that countries should open their borders. Caution is necessary while interpreting cross-sectional time series data when regression results are not completely robust to changes in the list of control variables. It is found in some empirical studies that tariff has positive correlation with growths in transitional and developing economies (Jozefina Semancikova, 2016, et al). More empirical study is necessary before any conclusion.

It is important to mention that any reforms, whether trade, economic policy or else, cannot be done in isolation and without any causal effect on other factors. Ceteris paribus does not work in real world setting. However, if it is possible to combine trade liberalization and adjustment to other important factors like governance, rule of law, anti-corruption, etc. into one combination, economic benefits can be reaped. Kyrre Stensnes, 2006 concluded that this benefit is twofold.

Even though trade openness and tariff have somewhat low explanatory interpretation on economic growth, significance shown in the results cannot be overruled. Lowering tariffs on imports is an important step stone of fighting against corruption as shown in Roberta Gatti, 1999. Tariffs can also lower the overall welfare of trade. Therefore, this can be said that international trade is an essential vessel for economic growth in transitional and developing countries with weak institutions and governance. Moreover, other sectors which contribute to economic growth can also benefit from increased international trade and lowered tariffs.

**Reference**

1. World Development Indicators, The World Bank Group.
2. https://www.rug.nl/ggdc/productivity/pwt/ for Penn World Table
3. Investopedia.com/comparativeadvantage
4. Azam Chaudhry, 2011. "Tariffs, Trade and Economic Growth in a Institutionals Quality," Lahore Journal of Economics, Department of Economics, The Lahore School of Economics.
5. https://taxfoundation.org/tariffs-trump-trade-war/, for the survey on Trump administration's trade war and tariff regime.
6. Aurangzeb, 2003. "Trade, Investment and Growth Nexus in Pakistan: An application of cointegration and multivariate causality test", vol 8, The Lahore Journal of Economics.
7. Romain Wacziarg, Karen Horn Welch, 2008; "Trade Liberalization and Growth: New Evidence", The World bank Economic Review, Volume 22, Issue 2.
8. Ghoshal I, 2015. "Trade-Growth relationship in India in the Pre and Post Trade Agreement regime," Procedia Economics and Finance, Elsevier.
9. Md. Abu Hasan, Md. Sanaullah, Mir Khaled Iqbal Chowdhury, Anita Zaman, 2017. "Trade-led Growth and Growth-led Trade Hypotheses in Bangladesh: An ARDLBounds Test Approach," Journal Economics, Management and Trade.
10. Anthony P. Thirlwall, 2000. "Trade, Trade Liberalisation and EconomicGrowth: Theory and Evidence", Economic Research Paper, African Development Bank.
11. Dan Ben David, Micheal B Loewy, 1998, "Free Trade, Growth and Convergence." Journal of Economic Growth.
12. Qazi Adnan Hye, Shahida Wizarat, Wee-Yeap Lau, 2013. "Trade-led growth hypothesis: An empirical analysis of South Asian Countries." Economic Modeling, volum 35.
13. Scitovski, scott, Balassa, 1993. "Openness, Trade Liberalization and Growth in Developing Countries", Journal of Economic Literature.

14. G20 Information Centre, www.g20.org.
15. Gene M. Grossman, Elhanan Helpman, 1994. "Endogenous Innovation in the Theory of Growth" Volum 8, Journal of Economic Perspective.
16. Nathan Nunn, Daniel Trefler, 2010. "The Structure of Tariffs and Long-Term Growth", Volume 2, American Economic Journal: Macroeconomics.
17. Faruq, A. T. M. O. (2023). The Determinants of Foreign Direct Investment (FDI): A Panel Data Analysis for the Emerging Asian Economies. European Journal of Business and Management Research, 8(6), 35–43. https://doi.org/10.24018/ejbmr.2023.8.6.2168
18. Janhavi Shanker Tripathi, 2016. "Trade Growth Nexus: A Study of G20 Countries", Volume 7, Journal of Economics and Finance.
19. Chia Yee Ee, 2015. "Export-Led Growth Hypothesis: Empirical Evidence from Selected Sub-Saharan African Countries", 7th International Economics and Business Management Conference, 5th & 6th October 2015, Science Direct.
20. Lewis E. Hill, Betsy Jane Clary, 1990; "Adam Smith on colonies: An analytical and historical interpretation", Forum for Social Economics, obtained from https://link.springer.com/article/10.1007/BF02761438
21. Yanikkaya, H., 2003; "Trade openness and economic growth: A cross-country empirical investigation", Journal of Development Economics, 72(2003)
22. Mohamed Fenira, 2015; "Trade Openness and Growth in Developing Countries: An Analysis of the Relationship after Comparing Trade Indicators", Asian Economic and Financial Review.
23. Robert E. Baldwin, 2004; "Openness and Growth: What's the Empirical Relationship?" http://www.nber.org/chapters/c9548
24. Jozefina Semancikova, 2016; "Trade, Trade openness and macroeconomic performance", 19th International Conference Enterprise and Competitive Environmen 2016, obtained from www.sciencedirect.com
25. Sergio Rebelo, 1991; "Long-Run Policy Analysis and Long-Run Growth" Journal of Political Economy.
26. Sabina Silajdzic, Eldin Mehic, 2018; "Trade Openness and Economic Growth: Empirical Evidence from Transition Economies", obtained from www.intechopen.com/books/ trade-and- global-market/trade-openness-and-economic-growth-empirical-evidence-from-transition- economies.
27. Dang Van Dan, Vu Duc Binh, 2018; "The Effect of Macroeconomic Variables Economic Growth: A Cross-Country Study", obtained from internet. Web address: link.springer.com/chapter/10.1007/978-3-030-04200-4_67
28. Yanqing Jiang, 2014; "4 - Potential effects of foreign trade on development ",
29. Elhanan Helpman, 2004; "The Mystery of Economic Growth", Bibliovault OAI Repository, the University of Chicago; abstract obtained from www.researchgate.net
30. David N DeJong, Marla Ripol, 2006; "Tariffs and Growth: An Empirical Exploration of Contingent Relationships." Review of Economics and Statistics, Volume 88, Issue 4, November 2006
31. Central Intelligence Agency (CIA) World Factbook. https://www.cia.gov/library/publications/the-world-factbook/

32. Kyrre Stensnes, 2006; "Trade Openness and Economic Growth: Do Institutions Matter?" Norwegian Institute of International Affairs, No. 702.
33. Daniel Hoechle; "Robust Standard Errors for Panel Regressions with Cross-Sectional Dependence", http://fmwww.bc.edu/repec/bocode/x/xtscc_paper.pdf
34. Roberta Gatti, 1999: "Corruption and Trade Tariffs, or a Case for Uniform Tariffs", World bank e-library https://elibrary.worldbank.org/doi/abs/10.1596/1813-9450-2216

**Appendix:**

1. The following figure is a test result on the necessity of time fixed-effects model. The result described above shows no requirement of time fixed effects as null hypothesis cannot be rejected.

```
testparm i.year

 ( 1)  1991.year = 0
 ( 2)  1992.year = 0
 ( 3)  1993.year = 0
 ( 4)  1994.year = 0
 ( 5)  1995.year = 0
 ( 6)  1996.year = 0
 ( 7)  1997.year = 0
 ( 8)  1998.year = 0
 ( 9)  1999.year = 0
 (10)  2000.year = 0
 (11)  2001.year = 0
 (12)  2002.year = 0
 (13)  2003.year = 0
 (14)  2004.year = 0
 (15)  2005.year = 0
 (16)  2006.year = 0
 (17)  2007.year = 0
 (18)  2008.year = 0
 (19)  2009.year = 0
 (20)  2010.year = 0
 (21)  2011.year = 0
 (22)  2012.year = 0
 (23)  2013.year = 0
 (24)  2014.year = 0
 (25)  2015.year = 0
 (26)  2016.year = 0
 (27)  2017.year = 0

       F( 27,   232) =    0.56
            Prob > F =  0.9644
```